\documentclass[11pt]{article}
\usepackage{graphics}
\usepackage{amssymb}
\usepackage{psfrag}
\usepackage{epsfig}
\usepackage{psfig}
\usepackage{epsf}
\usepackage{float}
\usepackage{latexsym}
\usepackage[german,USenglish]{babel}
\newcommand{\vs}{\vspace{-0.2cm}}
\newcommand{\beq}{\begin{equation}}
\newcommand{\eeq}{\end{equation}}
\newcommand{\beqa}{\begin{eqnarray}}
\newcommand{\eeqa}{\end{eqnarray}}

\newcommand{\bma}{\begin{array}{cc}}
\newcommand{\ema}{\end{array}}

\def\3{{\ss}}

\def\vek #1 {\overrightarrow {#1}}

\begin{document}

\hfill {\tiny FZJ-IKP(TH)-2002-19}

\vspace{1cm}

\begin{center}

{{\Large\bf Further comments on nuclear forces in the chiral limit}}

\end{center}

\vspace{.3in}

\begin{center}

{\large
E. Epelbaum,$^\dagger$\footnote{email:
                           evgeni.epelbaum@tp2.ruhr-uni-bochum.de}
Ulf-G. Mei{\ss}ner},$^\ddagger$\footnote{email:
                           u.meissner@fz-juelich.de}
W. Gl\"ockle$^\dagger$\footnote{email:
                           walter.gloeckle@tp2.ruhr-uni-bochum.de}

\bigskip

$^\dagger${\it Ruhr-Universit\"at Bochum, Institut f{\"u}r
  Theoretische Physik II,\\ D-44870 Bochum, Germany}\\

\bigskip

$^\ddagger${\it Forschungszentrum J\"ulich, Institut f\"ur Kernphysik
(Theorie),\\ D-52425 J\"ulich, Germany} \\
 and       \\
{\it Karl-Franzens-Universit\"at Graz, Institut f\"ur Theoretische Physik\\
A-8010  Graz, Austria}

\end{center}

\vspace{.3in}

\thispagestyle{empty}

\begin{abstract}
\noindent We substantiate our statement that the deuteron remains
bound in the chiral limit. We critically discuss recent claims that
effective field theory cannot give a definite answer to this question.
\end{abstract}

\noindent
\vfill

\pagebreak
\noindent {\bf 1.} The question of the quark mass dependence of the
nuclear forces has attracted considerable interest recently. The tool
to perform such investigations is chiral effective field theory (EFT), which
allows
for a unified description of pion--nucleon and nucleon--nucleon
dynamics. In \cite{EGM} we had shown that the deuteron remains bound
in the chiral limit, making use of an EFT approach based on a modified
Weinberg power counting which successfully describes two-, three- and
forur-nucleon systems. In earlier \cite{BBSvK} and parallel work
\cite{BS1} a different claim was made, namely that within the EFT one
presently cannot make a definite statement whether or not the deuteron is bound
in the chiral limit because some low-energy constants (LECs) are not
known.
In these works, a coordinate space power counting
was used which is supposedly equivalent to the one employed by us. In
a more recent paper, Beane and Savage \cite{BS2} tried to substantiate
their claim by performing an error analysis for the various input
parameters similar to what was done
already in \cite{EGM}. Here, we wish to critically examine various
claims made in that paper, and as a consequence, we are
able to {\it substantiate our original statement}.

\bigskip

\noindent {\bf 2.} Since we do not wish to repeat any of the detailed
arguments underlying our calculation presented in \cite{EGM}, we
rather concentrate here on some issues raised in \cite{BS2} which
deserve a closer look.
\begin{itemize}
\item[1)]In Ref.~\cite{BS2}, the naive dimensional analysis for the quark
(pion) mass dependent operators $\bar{D}_{I} M_\pi^2$ (where $I$ specifies the
channel
under consideration) is repeated in terms of a parameter $\eta$. It is
  claimed that we use $\eta \leq 1/19$ in the $^3S_1$ channel. We are
puzzled by such a statement. Even though the four--nucleon LECs are
to some extent regulator and convention dependent, such effects should
largely cancel in ratios of the LECs projected onto partial waves. So
what we have really used for this relative strength is \cite{EGM}:
\beq
\label{eq1}
\eta_{^1S_0} \equiv {\bar{D}_{^1S_0} \, M_\pi^2 \over {C}_0^{^1S_0}}
\leq {1 \over 4.2}~, \quad \eta_{^3S_1}
 \equiv {\bar{D}_{^3S_1} \, M_\pi^2 \over {C}_0^{^3S_1}}
\leq {1 \over
  8.6}~,
\eeq
which is  more than a factor of two different to what is claimed in
  \cite{BS2}. The values of the LECs ${C}_0^{^1S_0}$ and 
${C}_0^{^3S_1}$ corresponding to the exponential regulator defined in 
\cite{EGMres} and  the cut--off $\Lambda=560$ used here 
and in \cite{EGM} read:
\beq
C_0^{^1S_0} = -0.143 \times 10^4\mbox{ GeV}^{-2}\,, \quad \quad  \quad \quad 
C_0^{^3S_1} = -0.146 \times 10^4\mbox{ GeV}^{-2}\,.
\eeq
The appropriate (not partial wave projected) LECs $C_S$ and $C_T$, which enters the 
underlying chiral Lagrangian \cite{EGMres}, take the values:
\beq
C_S = -0.979 \, F_\pi^{-2}\,, \quad \quad  \quad \quad 
C_T = -0.003 \, F_\pi^{-2}\,,
\eeq
where we have used the same notation (and normalization) as in 
Ref.~\cite{EGMres}.\footnote{Some of the values given here are slightly 
different from the corresponding ones in  \cite{EGMres} due to the 
different value of $g_{\pi N}$ used, see \cite{EGM}.}
While the constant $C_S$ is perfectly natural with $\alpha_{C_S} =-0.979$,
the LEC $C_T$ appears to be very small as a consequence of the approximate Wigner symmetry,
see \cite{EGMres} for more details. The authors of \cite{BS2} should
provide better information on the size of their LECs in the
appropriate partial waves.
It is now important to address the question whether
the range for $\eta_I$ used in \cite{EGM} is too narrow? As we have explained
there, we have performed detailed and serious studies of the size
of LECs appearing at
NLO and NNLO in the chiral EFT and {\it none of
the LECs which can be determined from data exceeds $\pm 3$} when properly
  normalized in units of $F_\pi^2$ and $\Lambda_\chi^2$. Therefore,
the choice for $\eta_I$ made in \cite{EGM}, which corresponds to $-3 < \alpha_I < 3$, 
is well rooted in the phenomenology of four-nucleon
LECs. Also, if one were to extend this range to say $\pm 10$, one
would leave the realm of a converging EFT because NLO correction would
become comparable to the LO contributions. Thus, even as one claims to
know nothing about the quark mass dependent LECs, one should still
bound them based on such a convergence argument. Finally, as we have
discussed in \cite{EGM}, approximate Wigner symmetry requires small
values of the corresponding LECs. While such an argument is merely
aesthetic, it still should not be discarded completely.
\item[2)] The central result of Ref.~\cite{BS2} is Fig.4. This figure
as it is shown is more than confusing. First, all other figures are
scatter plots, so why are suddenly continuous lines drawn? Simply
connecting the most outside lying points by a line\footnote{As was
  explained to us by one of the authors.} makes not much sense.
In particular, the upper line explodes with decreasing pion
masses.  A simple extrapolation of this upper line to smaller values of
$M_\pi$ would probably lead to an unnaturally large deuteron
binding energy
of the order of few hundreds MeV and thus indicates the breakdown of the EFT.
This could indicate that the authors have considered a region
of parameter space that does not exclude deep bound states outside the
range of EFT.  The authors of \cite{BS2} do not clearly discuss 
the origin of such a dramatic
increase of the deuteron binding energy for  smaller values of $M_\pi$.
In Fig.~\ref{deut}, we show our results for the deuteron binding energy taking
the range of  $\bar{d}_{16}$ as given in the caption of Fig.~4 of \cite{BS2}
and increasing the value of $\eta_{^3S_1}$ given in eq.~(\ref{eq1}) and used 
in \cite{EGM} by a factor of two. The resulting value $\eta_{^3S_1}=1/4.3$
is even somewhat larger than the one suggested in \cite{BS2}.
The deuteron is bound in the chiral limit and the binding energy only varies
between 4 and 20 MeV.\footnote{We do not consider a variation in
  $\bar{d}_{18}$ as relevant here. The value we use refers to the
  well accepted small value of the pion--nucleon coupling constant as deduced
  by the
  Nijmegen and VPI/GWU groups. The larger uncertainty due to the fitting
  procedure in \cite{FM} should therefore be taken {\it cum grano salis}.
We will come back to that point later on.}
In our calculation, the expansion of the binding energy is well
behaved for all values of $M_\pi$. The scattering length in the $^3S_1$ partial 
wave calculated with the uncertainties in the LECs as suggested \cite{BS2} 
is depicted in Fig.~\ref{slen}. 
\begin{figure}[tb]
\centerline{
\psfig{file=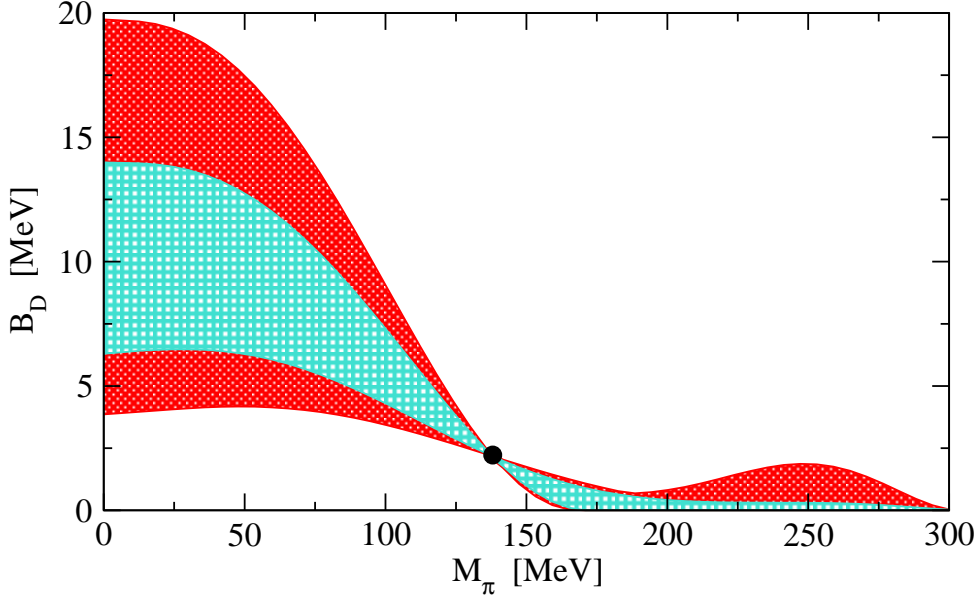,width=14cm}}
\vspace{0.3cm}
\centerline{\parbox{14cm}{
\caption[fig4]{\label{deut}
Deuteron binding energy as a function of the pion mass.
The shaded areas correspond to the allowed values.
The light shaded band refers to the variation
of  $\bar D_{^3S_1}$ using $\eta_{^3S_1} = 1/4.3$
and $\bar d_{16} = -1.23$ GeV$^{-2}$,  $\bar d_{18} = -0.97$ GeV$^{-2}$.
The dark shaded band gives the uncertainty if, in addition to variation of
$\bar D_{^3S_1}$, the LEC $\bar d_{16}$ is varied in the range from
$\bar d_{16} = -0.17$ GeV$^{-2}$ to $\bar d_{16} = -2.61$ GeV$^{-2}$.
The heavy dot shows the binding energy for the physical value of the
pion mass.
}}}
\vspace{0.7cm}
\end{figure}
\item[3)] There is an apparent inconsistency between Figs.~2,3 and
Fig.~4 of \cite{BS2}. If for pion masses below 60 MeV the deuteron can be
bound or unbound, the scattering length in the $^3S_1$ channel should
diverge as the deuteron passes from being bound to being unbound
provided the $M_\pi$--dependence of the deuteron binding energy is
continous. No
such divergence is seen in these figures. We also remark that one
rather should plot bands (as done in \cite{EGM}) because the variations
of these parameters cannot simply be treated as  statistical
fluctuations. Furthermore, as discussed in \cite{FBM}, there are
sizeable correlations between certain LECs in that analysis. It is
therefore highly questionable to use the MINUIT errors as given in
that paper for a statistical error analysis as done in \cite{BS2}.
\item[4)] It is not clear from Ref.~\cite{BS2} that the calculation is done
consistently with the power counting advocated in \cite{BBSvK}. It
that scheme one should treat the difference between the OPE exchange
and its chiral limit representation perturbatively. This is a very
slowly converging expansion for {\it small} energies as
demonstrated explicitly  in
\cite{BBSvK}.  As shown in Eq.~(3) of \cite{BS2}, the full OPE is used in
\cite{BS2}
and no uncertainty due to this procedure is given.
\item[5)] Keeping the short--range terms $\bar D_I M_\pi^2$ together with the
chiral limit representation of the two--pion exchange (TPE) in the potential
as is done in \cite{BS1}, \cite{BS2}
seems to be inconsistent. The reason is that the renormalized expression
for the leading chiral two--pion exchange contribution includes apart from
the non--polynomial pieces also terms of the kind \cite{EGM}
\beq
\label{cont_TPE}
\alpha_I  M_\pi^2 +
\beta_I  M_\pi^2 \ln \frac{ M_\pi}{\Lambda} \,,
\eeq
where $\alpha_I$ and $\beta_I$ are the known constants and $\Lambda$
refers to the renormalization scale.\footnote{Similar terms also arise
from renormalization of the short--range interactions by pion loops.}
\begin{figure}[tb]
\centerline{
\psfig{file=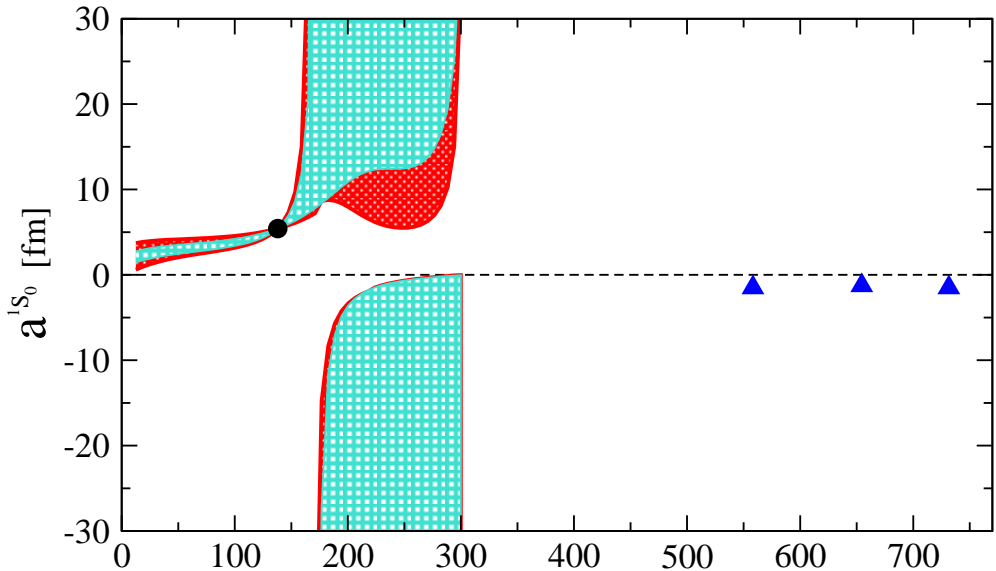,width=14cm}}
\vspace{0.3cm}
\centerline{\parbox{14cm}{
\caption[fig4]{\label{slen}
$^3S_1$--scattering length as a function of the pion mass.
The shaded areas correspond to the allowed values.
Bands as in Fig.~\ref{deut}. The heavy dot shows the scattering length 
for the physical value of the
pion mass. The triangles 
refer to lattice QCD results from \cite{Fukug95}. 
}}}
\vspace{0.7cm}
\end{figure}
This indicates that the $M_\pi$--dependence of the leading TPE is
as important as the $M_\pi$--dependence of the short--range terms
in eq.~(\ref{cont_TPE}). It is, therefore, not clear, why the authors
of \cite{BS1}, \cite{BS2} decided to neglect the explicit
$M_\pi$--dependence of the TPE as well as the second term in
 eq.~(\ref{cont_TPE}) and to keep only the first term in that equation.
\item[6)]The authors of \cite{BS2} claim to be able to reproduce our
  results using the same input parameters. As we just showed with
  respect to the discussion of their Fig.~4, we are not able to
  reproduce theirs. In particular, we obtain for the deuteron
  binding energy in the chiral limit: $B_D^{\rm CL}  = 9.6
  { {+ 4.4} \atop {-3.2}} { {+ 5.7} \atop {-2.4}}$ MeV, where the
  uncertainties for  $\bar{D}_{^3S_1}$ (the first error) and $\bar{d}_{16}$
  (the second error) are taken
  from \cite{BS2} or are even slightly larger, i.e.: $\eta_{^3S_1}  = 1/4.3$;
  $-2.61 \, \mbox{GeV}^{-2} < \bar{d}_{16} < -0.17\,  \mbox{GeV}^{-2}$.
  As already pointed out, we do not consider a variation in $\bar d_{18}$ as
  relevant here, and, therefore, have not plotted the corresponding bands in
  Fig.~\ref{deut}. For the sake of completeness, we however calculated the
  resulting additional uncertainty in the chiral limit value of the deuteron
 binding
  energy. Note that the LEC $\bar d_{18}$ does not contribute to the OPE
in the chiral
  limit (where the Goldberger-Treiman relation is exact)
and thus changing $\bar d_{18}$ only affects $B_D^{\rm CL}$ indirectly,
  due to corresponding small changes in the LECs related
to contact interactions,
  as explained in \cite{EGM}. Variation of  $\bar d_{18}$ in the range
  $-1.54  \, \mbox{GeV}^{-2}  <  \bar d_{18} < -0.51  \,
\mbox{GeV}^{-2}$ in addition to
  variation of $\bar{D}_{^3S_1}$ and $\bar{d}_{16}$ as suggested
  in \cite{BS2} leads to $B_D^{\rm CL}
= 9.6  { {+ 4.4} \atop {-3.2}} { {+ 5.7} \atop {-2.4}}
  { {+ 0.6} \atop {-0.4}}$ MeV. As expected, the
resulting additional uncertainty is quite
  small.
\item[7)]Some of parameter choices discussed in \cite{BS2} appear
  questionable. While all fits of  $\bar{d}_{16}$ in \cite{FBM}
  give negative values for this LEC, the authors of \cite{BS2} claim
  that a positive value of $\bar{d}_{16} = +1\,$GeV$^{-2}$ is
  consistent with the data. It is their duty to present an analysis of the
many data on $\pi N \to \pi\pi N$ using this particular value for
$\bar{d}_{16}$. It should also be stressed that the various fits
of $\pi N \to \pi \pi N$ in \cite{FBM} are for exactly {\it one}
choice of $\pi N$ phases and thus {\it one} value for $\bar{d}_{18}$
(the phase shifts of \cite{matsinos}).
Therefore, the independent variation of $\bar{d}_{16}$ and
  $\bar{d}_{18}$, the latter referring to other $\pi N$ phase
shift analyses, as done in Ref.~\cite{BS2} does not really make sense.
\end{itemize}

\bigskip
\noindent {\bf 3.} In summary, we have shown that the results
presented in \cite{EGM} are robust under parameter variations,
in particular, the deuteron {\em is bound in the chiral limit}.
We have also shown that some of the results presented in \cite{BS2}
cannot be understood simply and thus need clarification.
We stress again that the discussions presented here do not influence any of the
statements made in our previous work \cite{EGM},
all results and conclusions of that paper remain unchanged.

\vspace{1cm}
\noindent {\bf Acknowledgements}
We thank Silas Beane for some clarifying discussions.

\end{document}